\documentclass[11pt]{article}
\usepackage[utf8]{inputenc}
\usepackage{amsmath, amsfonts, amssymb}
\usepackage[a4paper, margin=1in]{geometry}
\usepackage{setspace}
\usepackage{newpxtext}
\usepackage{authblk}
\usepackage{graphicx}

\usepackage[hyperindex,breaklinks]{hyperref}
\usepackage{xcolor}
\usepackage{booktabs}
\usepackage[group-separator={,}]{siunitx}
\usepackage{float}
\usepackage{wrapfig}

\title{Oil \& Water?\\ Diffusion of AI Within and Across Scientific Fields}

\author[1,4]{\small Eamon Duede\thanks{Correspondence to: eduede@g.harvard.edu}}
\author[2,5]{\small William Dolan}
\author[2,5]{\small Andr\'e Bauer}
\author[2,3,5]{\small Ian Foster}
\author[1,4]{\small Karim Lakhani}
\affil[1]{Harvard University}
\affil[2]{University of Chicago} \affil[3]{Argonne National Laboratory}
\affil[4]{Digital Data Design Institute}
\affil[5]{Globus Labs}

\date{} % Optional

\begin{document}

\maketitle
\begin{abstract}

This study empirically investigates claims of the increasing ubiquity of artificial intelligence (AI) within roughly 80 million research publications across 20 diverse scientific fields, by examining the change in scholarly engagement with AI from 1985 through 2022. We observe exponential growth, with AI-engaged publications increasing approximately thirteenfold (13x) across all fields, suggesting a dramatic shift from niche to mainstream. Moreover, we provide the first empirical examination of the distribution of AI-engaged publications across publication venues within individual fields, with results that reveal a broadening of AI engagement within disciplines. While this broadening engagement suggests a move toward greater disciplinary integration in every field, increased ubiquity is associated with a semantic tension between AI-engaged research and more traditional disciplinary research. Through an analysis of tens of millions of document embeddings, we observe a complex interplay between AI-engaged and non-AI-engaged research within and across fields, suggesting that increasing ubiquity is something of an oil-and-water phenomenon---AI-engaged work is spreading out over fields, but not mixing well with non-AI-engaged work.

\end{abstract}

\section{Introduction}

Interest in artificial intelligence (AI) has grown rapidly in recent years due, in large part, to fundamental breakthroughs \cite{vaswani2017attention, ouyang2022training, rafailov2023direct,child2019generating} that hold the promise to transform science and society. This promise has led to an extraordinary increase in AI-related research output in fields such as Computer Science that have traditionally been principally responsible for the development of, and innovation upon, emerging AI technologies \cite{krenn2023forecasting}. Moreover, applications of these breakthroughs have led to substantive impacts on scientific questions in pure and applied mathematics \cite{davies2021advancing,romera2023mathematical, han2018solving, duede2024apriori}, molecular biology \cite{jumper2021highly}, quantum chemistry \cite{gomez2016design}, physics \cite{iten2020discovering}, materials science \cite{tshitoyan2019unsupervised}, and even social and cultural sciences, the humanities \cite{kozlowski2019geometry,duede2022instruments,duede2024humanities}, and business \cite{dell2023navigating}.

However, the broad visibility of particularly prominent fundamental and applied breakthroughs tells us little about the true extent of engagement with AI across science and scholarship. While broad enthusiasm and a steady cadence of discoveries suggest that AI research is no longer a narrow specialization within a small number of fields with well-defined applications, but rather is becoming a more general and ubiquitous element of scholarship \cite{zenil2023future}, surprisingly little is known about how AI is reshaping diverse fields and the extent to which it is becoming an integral thread in the fabric of scientific inquiry and societal development.
Thus, in this paper, we step back and empirically evaluate whether and to what extent AI is, in fact, becoming ubiquitous in science and scholarship. 

We note that, while others have observed a rapid rise in AI engagement \cite{krenn2023forecasting,klinger2021deep,frank2019evolution,ben_blaiszik_2023_7713954}, increasing engagement is not necessarily an indication of increasing ubiquity. In this context, `ubiquity' refers to the extent of widespread engagement with AI within and across various subfields, rather than concentrated engagement by a small proportion of scientists or groups within a field. Across 20 scientific and scholastic fields, we examine the rise of AI engagement within roughly 80 million papers and evaluate the extent to which engagement with AI is becoming 'ubiquitous' within and across them. That is, we explicitly measure and analyze field-level penetration of AI-engaged scholarship. We do this by systematically investigating three related research questions (\textbf{RQs}).

\textbf{RQ1: How has the fraction of scientific and scholastic research that engages with AI changed over time?} In answering this question, we observe changes in the percentage of AI-engaged publications in each of 20 academic fields from 1985 through 2022. Perhaps unsurprisingly, we observe rapid growth in AI engagement across all fields. While this growth is suggestive of broadening ubiquity, the observed increase in AI engagement could be explained either by dynamics that resemble bubbles or by dynamics that resemble broader diffusion, where only the latter is characteristic of increasing ubiquity. This distinction leads naturally to our second research question.

\textbf{RQ2: To what extent has AI engagement diffused within individual fields?} To answer this question, we measure changes in the distribution over time of AI-engaged research across publication venues within each of the 20 fields. In general, we find that, even within fields in which penetration of AI engagement remains limited, the rate of diffusion is rapidly increasing. This suggests that the last several years have the beginning of a rapid broadening of AI engagement within scientific and scholastic disciplines. These findings inspire our third research question.

\textbf{RQ3: How is changing ubiquity of AI engagement associated with changes in the semantic character of fields?} In answering this question, we study how changes in the ubiquity of AI engagement within a field are associated with changes in the themes, topics, and focus areas of research papers that do and do not engage with AI. In particular, we measure the association between changes in the semantic characteristics of a field's AI-engaged papers and that of its Non-engaged papers relative to AI-engaged papers published in Computer Science over time. We find that increasing ubiquity within a field is associated with that field's AI-engaged papers becoming \textit{more} semantically similar to the AI-engaged papers of Computer Science and \textit{less} similar to all other research published in the field. We argue that these shifts in trajectories should be viewed as evidence of a kind of semantic tension within fields which, despite increasing ubiquity, could well be making it difficult for AI-engaged research to mix well with more `traditional' research. 

Our results provide the first field-level, quantitative evaluation of increasing ubiquity of AI engagement over time, as well as the first evaluation of how change in ubiquity is associated with changes in the semantics of scientific and scholastic disciplines.

\section{Data}
\label{sec:data}

The results of this paper draw on analyses performed on the Semantic Scholar \texttt{Academic Graph} \cite{kinney2023semantic}, a massive, open access index of scholarly literature containing rich metadata on hundreds of millions of academic articles and books. To start, we downloaded and processed the full \texttt{papers} (e.g., paper metadata), \texttt{abstracts}, \texttt{citations}, and document vector \texttt{embeddings} datasets from the 2023-08-01 release of the complete \texttt{Academic Graph}\footnote{The \texttt{Academic Graph} dataset is available at \url{https://api.semanticscholar.org/api-docs/datasets}.}. Entries in these datasets are cross-linked via the unique corpus identifiers that are assigned to paper metadata entries. The downloaded corpora included 213,600,818 paper metadata entries, 101,284,402 abstracts, and 134,705,300 document embeddings. 
The varying counts across datasets reflect the varying degrees of completeness in the information recorded in \texttt{Academic Graph} for different papers:
specifically, while every abstract and every embedding corresponds to a paper, not every paper has a corresponding abstract or an embedding.

Of particular importance for this study is the Semantic Scholar paper metadata dataset which classifies papers as belonging to one of 23 distinct scholarly fields. This classification is performed by Semantic Scholar using the \texttt{S2FOS}\footnote{The code, data, and model for \texttt{S2FOS} are available at \url{https://github.com/allenai/s2\_fos}} classifier, which recapitulates the \texttt{Microsoft Academic Graph} field-level classifications but expands coverage to include four additional fields: `Linguistics,' `Law,' `Education,' and `Agriculture and Food Sciences.' The paper metadata also assigns each paper to a publication venue (\textbf{RQ2}) in which it appears. See \cite{kinney2023semantic} for a description of how Semantic Scholar standardizes and normalizes publication venues.

% The analysis we present in what follows is conducted on subsets of the full data
% We analyze a subset of the \texttt{papers} corpus where we have classified AI engagement through the presence of related keywords in their abstracts (see Section~\ref{sec:rq1_methods}). 

We then process each of the downloaded datasets into subsets for analysis. To create our dataset of papers for analysis (\textbf{RQ1} and \textbf{RQ2}), we implemented a number of pre-processing steps on the full \texttt{Academic Graph}. First, we removed all non-English language texts from the  \texttt{abstracts} dataset using Meta's \texttt{fastText} model \cite{joulin2016bag}. We then cleaned the remaining abstracts by removing nonalpha characters and common stopwords, as well as lemmatizing individual words. The language-filtering and cleaning properly formatted the remaining text for keyword classification. After classifying abstracts as engaged or not engaged with AI, all were subsequently associated with their corresponding paper metadata entries for further analysis via their shared corpus ids, thus creating our dataset \texttt{classified-papers}. 

For our analysis of the changes in the semantics of fields (\textbf{RQ3}), we created an additional dataset of the document embeddings provided in the full \texttt{embeddings} dataset from the \textbf{Academic Graph}. Document embedding vectors are numerical representations of documents that capture their semantic content in a high-dimensional space. Embeddings were generated by using the \texttt{SPECTER2} \cite{cohan2020specter} model, which is fine-tuned on \texttt{SciBERT} \cite{beltagy-etal-2019-scibert}. \texttt{SPECTER2} processes paper titles and abstracts, and optimizes its performance by reducing a triplet margin loss \cite{cohan2020specter}. The aim is to promote learning of more closely aligned embeddings for pairs of papers that cite each other, compared to those pairs that do not share a citation relationship.\footnote{Code, data, and \texttt{SPECTER} model are available at \url{https://huggingface.co/allenai/specter2}}. We create our \texttt{classified-embeddings} dataset by associating vectors from the full Semantic Scholar embeddings dataset with metadata and classifications from our \texttt{classified-papers} dataset via their shared corpus id.

Given that today's AI technologies are significantly different both technologically and theoretically from those employed in earlier research eras, we restrict our analysis to papers that appear from 1985 onwards. This timing aligns with the historical shift in the mid-1980s away from broadly symbolic systems and toward the early connectionist approaches that dominate AI research today. Finally, excluding data from 2023, as that year is incomplete, we thus focus on 1985 through 2022. Moreover, due to uneven coverage year over year in the \texttt{Academic Graph}, we exclude the fields and associated papers from `law,' `sociology,' and `geography' from our analysis. This results in our final analysis datasets \texttt{classified-papers-recent} and \texttt{classified-embeddings-recent}. All results presented in Section~\ref{sec:class_findings} are produced from analyses carried out on \texttt{classified-papers-recent} and \texttt{classified-embeddings-recent}. See Table~\ref{tab:datasets} for a breakdown of the datasets we synthesized from the Semantic Scholar corpus and how these were used in the present study.

\begin{table}[!ht]
\centering
\label{tab:datasets}
\begin{tabular}{lclclc|}
\hline
 \textbf{Dataset} & \textbf{Number of Entries} & \textbf{Use} \\
 \hline
\texttt{english-abstracts} & 83,968,032 & Cleaning \\
\texttt{english-cleaned-abstracts} & 83,858,929 & Keyword classification \\
\texttt{classified-embeddings} & 83,327,504  & Ubiquity analysis \\
\texttt{classified-abstracts} & 83,858,929 & Combined with paper metadata \\
\texttt{classified-papers}   & 83,453,690 & Engagement tracking\\
\texttt{\textbf{classified-papers-recent}} &  77,399,492 & Engagement analysis (\textbf{RQ1}, \textbf{RQ2})\\

\texttt{\textbf{classified-embeddings-recent}} & 77,276,518  & Semantic analysis (\textbf{RQ3})\\
 \hline
\end{tabular}
\caption{Datasets synthesized from Semantic Scholar corpora}
\end{table}

\section{Methods and Findings}

\subsection{RQ1: How has the percentage of scientific and scholastic research that engages with AI changed over time?}

\subsubsection{Methods}
\label{sec:rq1_methods}

We seek to identify all papers that are engaged with AI, regardless of how they engage. Engagement could include (but is not limited to) the development of novel AI theory and approaches, technologies, or applications; the general use of AI models for domain-specific tasks; and critical engagement with AI, as typified by academic discourse in fields like philosophy and ethics. Thus, ours is a binary classification task: a paper is either engaged (henceforth `AI-engaged') or not engaged (henceforth `Non-AI-engaged'). To accomplish this task, we develop a keyword-based approach to classifying documents.

\textbf{Keyword Identification}: In developing our classifier, we elected to focus on identifying papers that are engaged with AI as it is currently conceived. It in important to note, however, that there have been several broad `waves' of AI research, each focused on an otherwise heterogeneous assemblage of technological approaches \cite{dreyfus1995making,mitchell2019artificial}. For instance, early waves in the mid-20th century through the 1980s (now broadly synonymous with the `Good old fashioned AI,' or GOFAI, period) focused on symbolic reasoning, logic, and the attempt to replicate human intelligence through explicit, rule-based systems \cite{feigenbaum1963computers, russell2010artificial}. The vocabulary characteristic of AI research in that period is highly distinct from that typified by the so-called `connectionist' approaches out of which most contemporary architectures (e.g., deep neural networks) emerged.

To develop a comprehensive list of keywords focused on contemporary approaches to AI, we began by generating a small `seed list' $K_s$ of obvious key terms (e.g., ``artificial intelligence,'' ``deep neural network,'' ``machine learning''). We then took the following steps to iteratively expand and refine our keyword list. 

First \textbf{(1)}, we downloaded all papers from the arXiv.org \texttt{cs.AI}, \texttt{cs.LG}, \texttt{cs.NE}, and \texttt{stat.ML} categories which, collectively, provide a fairly representative corpus of full-text documents that are engaged with AI in some way. Next \textbf{(2)}, we used the \texttt{Gensim} implementation\footnote{The implementation is available at \url{https://radimrehurek.com/gensim/models/word2vec.html}} of the popular word embedding algorithm \texttt{word2vec} \cite{mikolov2013efficient} to generate vector representations for each word in the arXiv corpus vocabulary, following standard pre-processing approaches. Then \textbf{(3)}, for each term $k_{si}$ in our initial `seed list,' we determined the 10 terms $t_i \in T$ with the highest cosine similarity (see Equation~\ref{eq:similarity}) to $k_{si}$ in the generated vector space and included them in a new `augmented list' $K_a = K_s \cup T$. Next \textbf{(4)}, we asked active AI experts at the University of Chicago, Harvard University, and Argonne National Laboratory to evaluate $K_a$ with an eye toward removing terms that could trigger false positives\footnote{For instance, terms like `hyperparameter' and `convolutional' have statistical and mathematical uses that are wholly unrelated to artificial intelligence. Additionally, certain abbreviations like `ML' were excluded as they have meanings (e.g., milliliter) that are unrelated to AI.}, as well as adding terms that pick out AI engagement with certainty. Finally \textbf{(5)}, we repeated steps 3 and 4 twice more to arrive at a final, expert-evaluated, keyword list $K_f$.

\textbf{Classification}:  To classify papers as either AI-engaged or Non-AI-engaged, we iterated through the abstracts of each paper and labelled a paper as AI-engaged if the paper abstract contains at least one term from $K_f$. 
\subsubsection{Findings}
\label{sec:class_findings}

\begin{figure}[!ht]
\centering
\includegraphics[width=.66\linewidth]{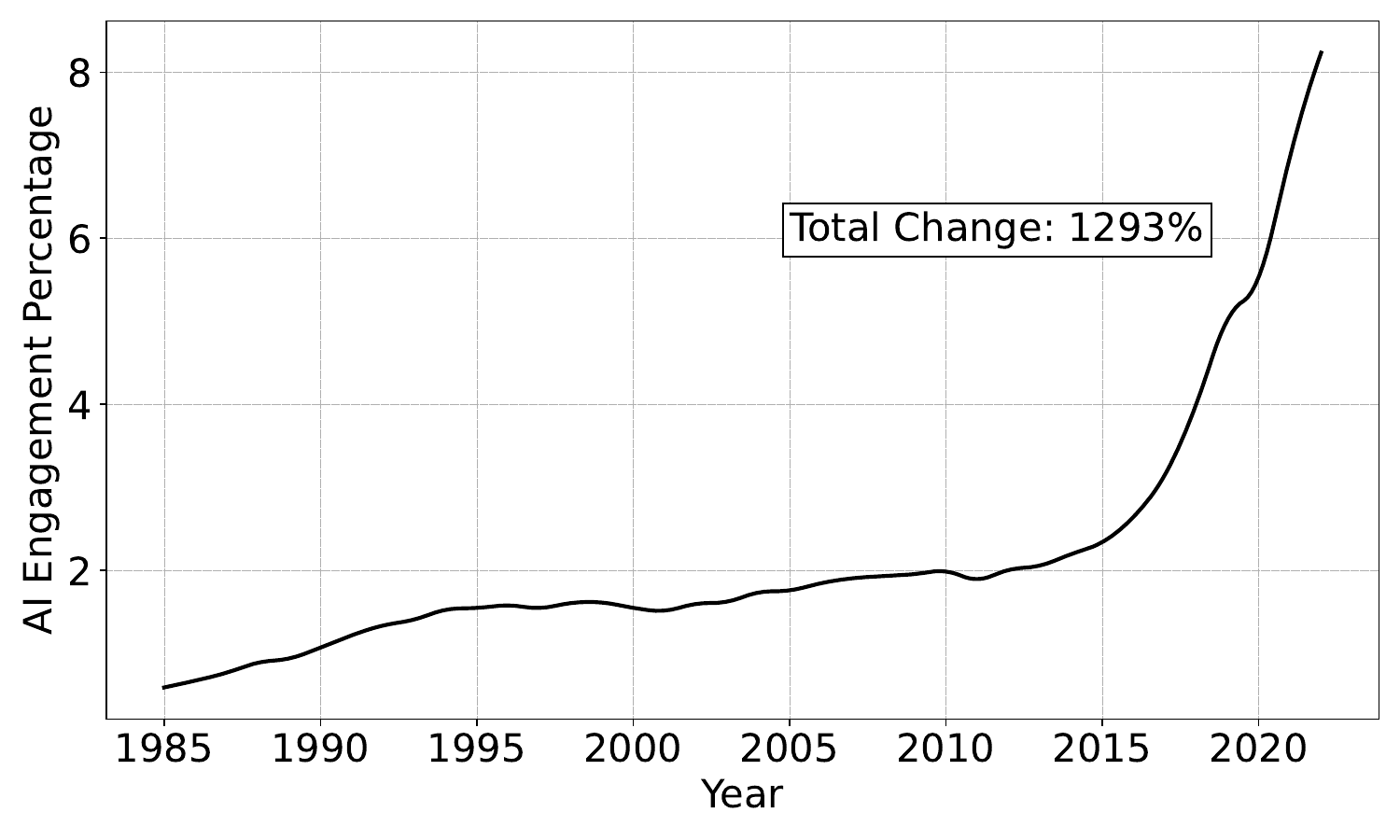}
\caption{Change in AI engagement across all fields from 1985 - 2022}
\label{fig:total_growth}
\end{figure}

In 2022, nearly 9\% of all papers in our \texttt{classified-papers-recent} dataset were identified as AI-engaged. Figure~\ref{fig:total_growth} shows the change in the percentage of papers classified as AI-engaged over time. Over the period from 1985 through 2022, all fields represented in our \texttt{classified-papers-recent} corpus have experienced collective growth in AI engagement of 1293\%.

\begin{figure}[!ht]
\centering
\includegraphics[width=0.9\linewidth]{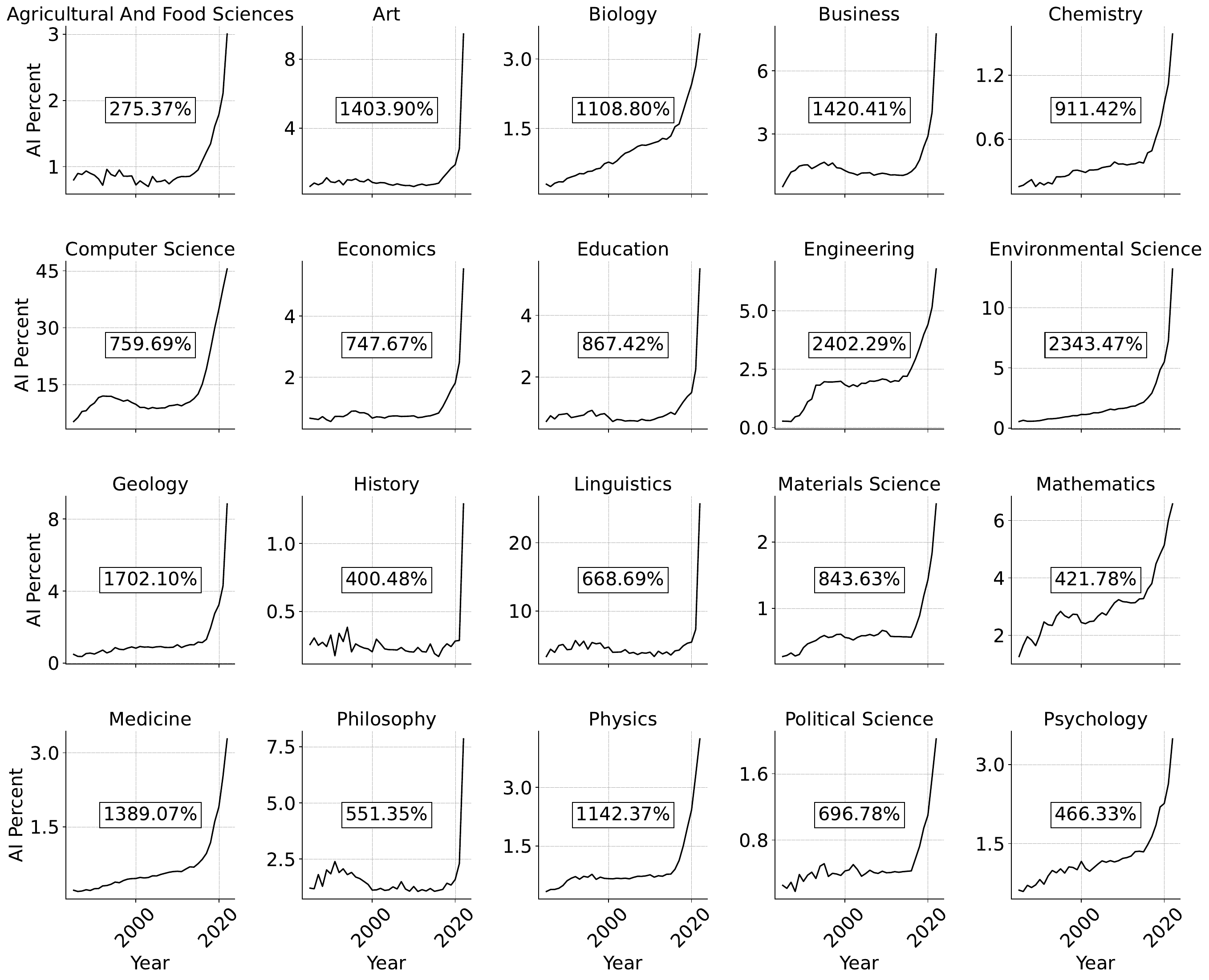}
\caption{Change in AI engagement percentage from 1985 - 2023 by field. Inserts tally the total change in percentage of AI-engaged publications for each field.}
\label{fig:field_growth}
\end{figure}

At the level of individual fields, we observe strikingly rapid growth in AI engagement across all of science. Figure~\ref{fig:field_growth} represents change from 1985 - 2022 in the percentage of AI-engaged papers in each of the 20 fields that we analyzed.

These findings lend robust empirical support to the already widely held belief that engagement with AI is increasing rapidly throughout science. We note that these trends are observed in the scholarly literature prior to the release of OpenAI's ChatGPT, suggesting that broad research engagement was already rapidly accelerating before the emergence of the most powerful generative large language models.

\subsection{RQ2: To what extent has AI engagement diffused within individual fields?}

\subsubsection{Methods}
\label{sec:methods}
While we observe a rapid increase in AI-engaged publications across all fields, that increase might be due to dynamics that are unrelated to increasing ubiquity. For instance, it could be that, within fields, new publication venues (e.g., journals and conferences) have emerged and are absorbing the bulk of new AI-engaged work. If this were the case, then the observed pattern of increasing AI engagement would be explained by bubble-like dynamics within fields, rather than by increasing ubiquity. To evaluate the extent to which AI engagement can be said to be increasingly ubiquitous, we directly measure the concentration of AI-engaged publications across a field's publication venues.

To evaluate ubiquity, we measure how concentrated AI-engaged papers are within the publication venues of individual fields by calculating the Gini coefficient of the observed distribution of AI-engaged papers across all journals within each field. The Gini coefficient is widely used in social scientific research to measure dispersion in frequency distributions \cite{mcdonald2008generalized} and has recently been used to observe the concentration of training datasets across AI research \cite{koch2021reduced}. Gini ranges continuously between 0 and 1, with, in the present case, 0 indicating that AI-engaged papers within a field are distributed across all publication venues in equal proportions, while 1 indicates that a single publication venue is responsible for publishing all of a field's AI-engaged papers.

To calculate Gini we generate, for each year and for each field, an array of length $n$ where $n$ is the number of publication venues in that field for that year. All elements $X$ in the array are sorted in ascending order such that $x_i$ represents the percentage of AI-engaged papers for the $i^{th}$ journal in the array. The Gini coefficient $G$ is then defined on this data as:
\begin{equation}
\label{eq:gini}
G=\frac{\sum_{i=1}^n(2 i-n-1) x_i}{n \sum_{i=1}^n x_i}.
\end{equation}

Given that $G$ ranges from 0 to 1, where lower values indicate lower inequality, we define \texttt{Ubiquity} as:
\begin{equation}
\label{eq:ubiquity}
    \text{\texttt{Ubiquity}} = 1 - \text{G},
\end{equation}

\noindent
whereby increasing ubiquity is more intuitively represented by values approaching 1. This approach gives us a precise measure of ubiquity for each field for each year. We then calculate the \texttt{Ubiquity} scores for all fields in all years.

\subsubsection{Findings}
\begin{figure}[!ht]
\centering
\includegraphics[width=0.5\linewidth]{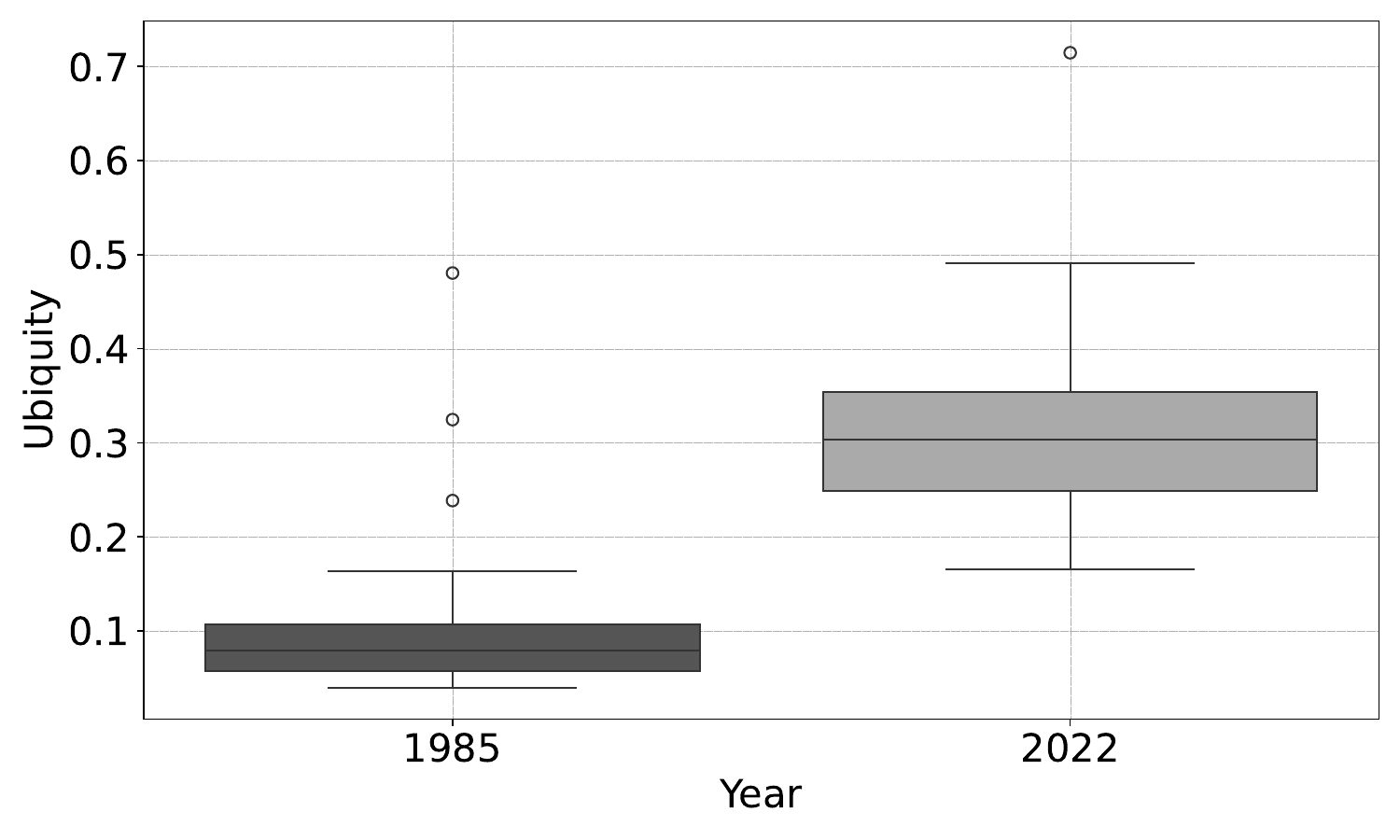}
\caption{Boxplots representing the overall change in \texttt{Ubiquity} from 1985 to 2023.}
\label{fig:ubiquity_shift}
\end{figure}

Relative to 1985, \texttt{Ubiquity} has risen substantially. Figure~\ref{fig:ubiquity_shift} shows the change in the distribution of \texttt{Ubiquity} across fields from 1985 to 2023.

\begin{figure}[!ht]
\centering
\includegraphics[width=.9\linewidth]{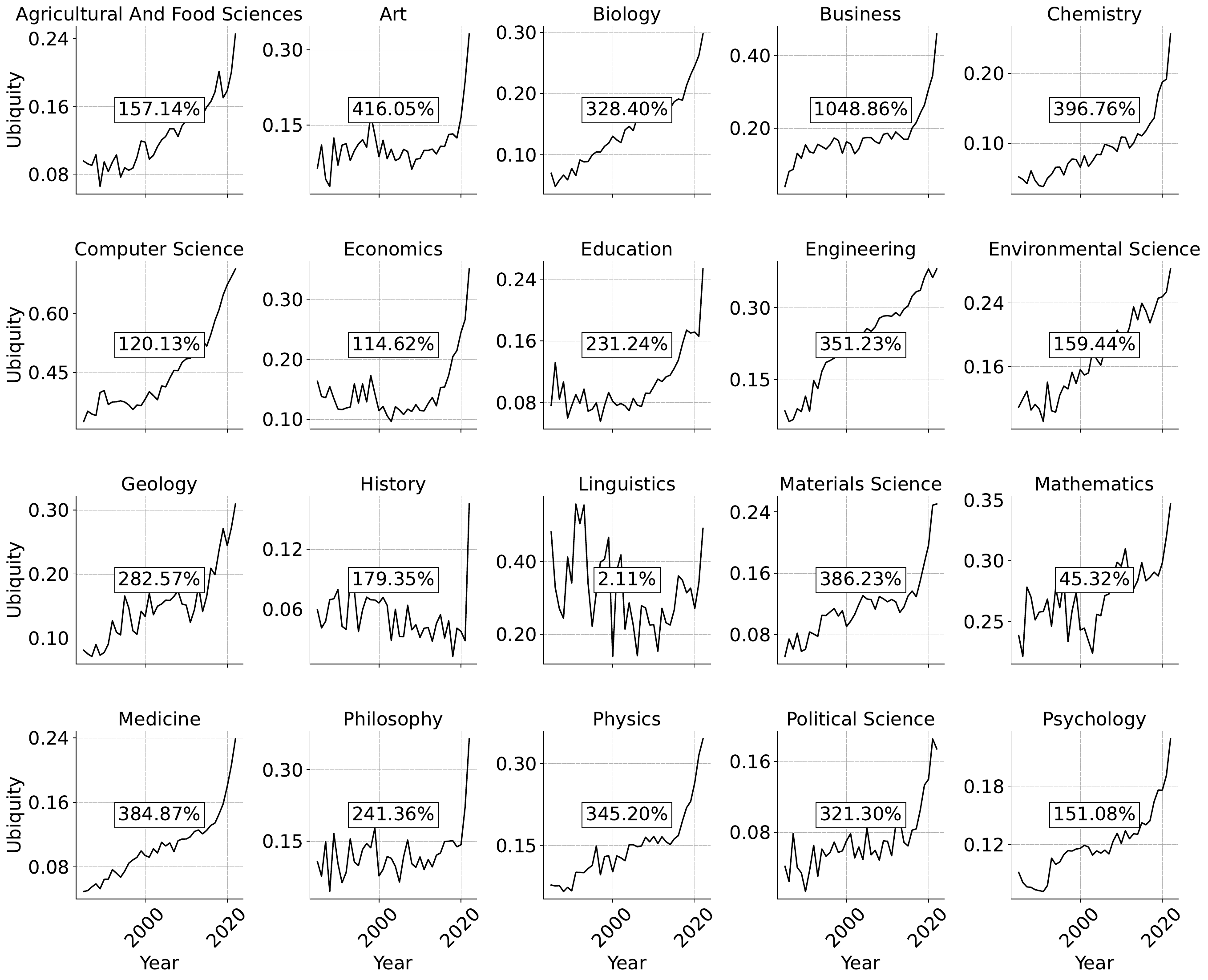}
\caption{\texttt{Ubiquity} over time from 1985 - 2022. Inserts tally the total percentage change in Ubiquity for each field.}
\label{fig:field_ubiquity}
\end{figure}

We observe that the trajectory of the rate at which AI engagement is becoming more ubiquitous within fields in the last decade is striking (see Figure~\ref{fig:field_ubiquity}) with every field in our corpus experiencing a rapid increase in the diffusion of AI-engaged research across their publication venues. 

These findings provide the first concrete, empirical detection and measure of changes in field-level ubiquitous engagement with AI in science and scholarship that we are aware of.

\subsection{RQ3: How is changing ubiquity of AI engagement associated with changes in the semantic character of fields?}

\subsubsection{Methods}
Finally, we ask whether and to what extent increasingly ubiquitous engagement with AI is associated with relative changes in the intellectual trajectory of fields captured by changes in the semantics (e.g., changes in themes, topics, focus areas, etc.) of papers. To measure changes in intellectual trajectory, we analyze changes over time in the relative distances between papers in our \texttt{classified-embeddings-recent} corpus represented in document embedding space. Document embeddings generated using the \texttt{SPECTER2} model represent rich encodings of each paper's semantics. In embedding spaces such as this, semantic distances between papers can be measured with precision and have been used widely to represent intellectual similarities and distances between and across publications \cite{kozlowski2019geometry,duede2024being,lewis2023local,aceves2024human,teplitskiy2022effect}. Here, we analyze how changes in \texttt{Ubiquity} are associated with changes in the semantics of scientific publications within fields.

First, we associate the AI engagement status, year, date, and field-level venue of papers in our corpus to their corresponding embeddings. Using this data, we obtain yearly centroids (mean vectors) for each field's sets of AI-engaged papers and Non-AI-engaged papers. These centroids represent the `average' semantic character of each fields' AI-engaged papers and Non-AI-engaged papers, allowing us to measure directly the semantic similarity between them.

Next, for each year, we calculate the semantic similarity (Equation~\ref{eq:similarity}) between field-level centroids for the AI-engaged and Non-AI-engaged sets \emph{A} from all fields and the AI-engaged centroid \emph{B} for Computer Science in that year. 

\begin{equation}
\label{eq:similarity}
\text{similarity}(A, B)=\frac{\sum_{i=1}^n A_i \times B_i}{\sqrt{\sum_{i=1}^n A_i^2} \times \sqrt{\sum_{i=1}^n B_i^2}}
\end{equation}

This measure allows us to observe how field-level papers (whether AI-engaged or not) are changing relative to Computer Science. The intuition here is that one way for papers to become more AI-engaged is for them to take on questions and problems in a manner that reflects the semantic character of AI papers in Computer Science. If this were the case, then, as AI engagement increases, we would observe an associated field-level shift toward Computer Science. However, another way for papers to become more AI-engaged is for them to take on questions concerned with AI in a manner that reflects the semantic character of their own discipline. If this were the case, then we would expect not to  observe a meaningful shift in a field's AI-engaged papers away from the its Non-AI-engaged papers, nor would we expect to observe a shift toward Computer Science. Finally, within each field, we measure the similarity between that field's own AI-engaged and Non-AI-engaged papers.

These measurements yield the following three variables: \texttt{ai\_similarity}, which measures the semantic similarity between the centroid of each field's AI-engaged papers and the centroid of the AI-engaged papers in Computer Science for each year; \texttt{non-ai\_similarity}, which measures the semantic similarity between the centroid of each field's Non-AI-engaged papers and the centroid of the AI-engaged papers in Computer Science for each year; and \texttt{inner\_similarity}, which measures the semantic similarity between the centroid of each field's AI-engaged and Non-AI-engaged papers.

\begin{figure}[!ht]
\centering
\includegraphics[width=0.7\linewidth]{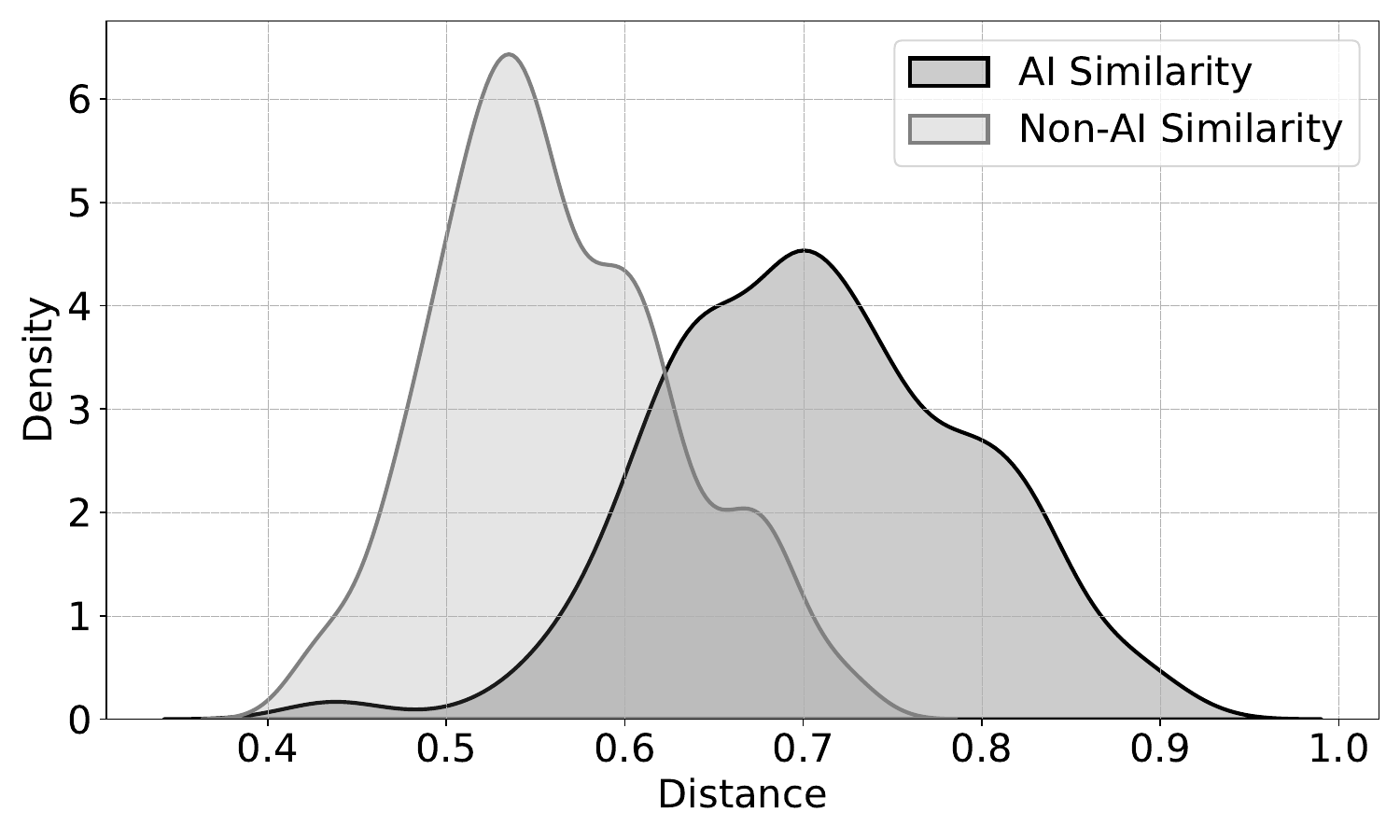}
\caption{Density of the distribution of semantic similarity between the centroids of all Non-AI-engaged and all AI-engaged papers and the centroids of AI-engaged papers in Computer Science across all fields for the period 1985 - 2023. Values closer to 1 on the x-axis represent higher semantic similarity.}
\label{fig:field_centroid}
\end{figure}

As is evident from Figure~\ref{fig:field_centroid} and Table~\ref{tab:summary_stats}, across all fields and years in our sample, AI-engaged papers are, in general, more semantically similar to the AI-engaged papers published in Computer Science than are Non-AI-engaged papers. This result should not come as a surprise, as AI-engaged papers across fields engage with many of the same themes and thus will evince similar semantic characteristics. However, across all fields in our sample the semantic similarity between AI-engaged and Non-AI-engaged papers within fields (\texttt{inner\_similarity}) is quite high, which is also to be expected. After all, in order for papers to be published within a given discipline, they have to be recognizable as contributions to that discipline \cite{shi2023surprising,duede2021social}. Importantly, however, in the period 1985 - 2020, \texttt{inner\_similarity} has decreased in all but 4 fields. 

\begin{figure}[!ht]
\centering
\includegraphics[width=0.7\linewidth]{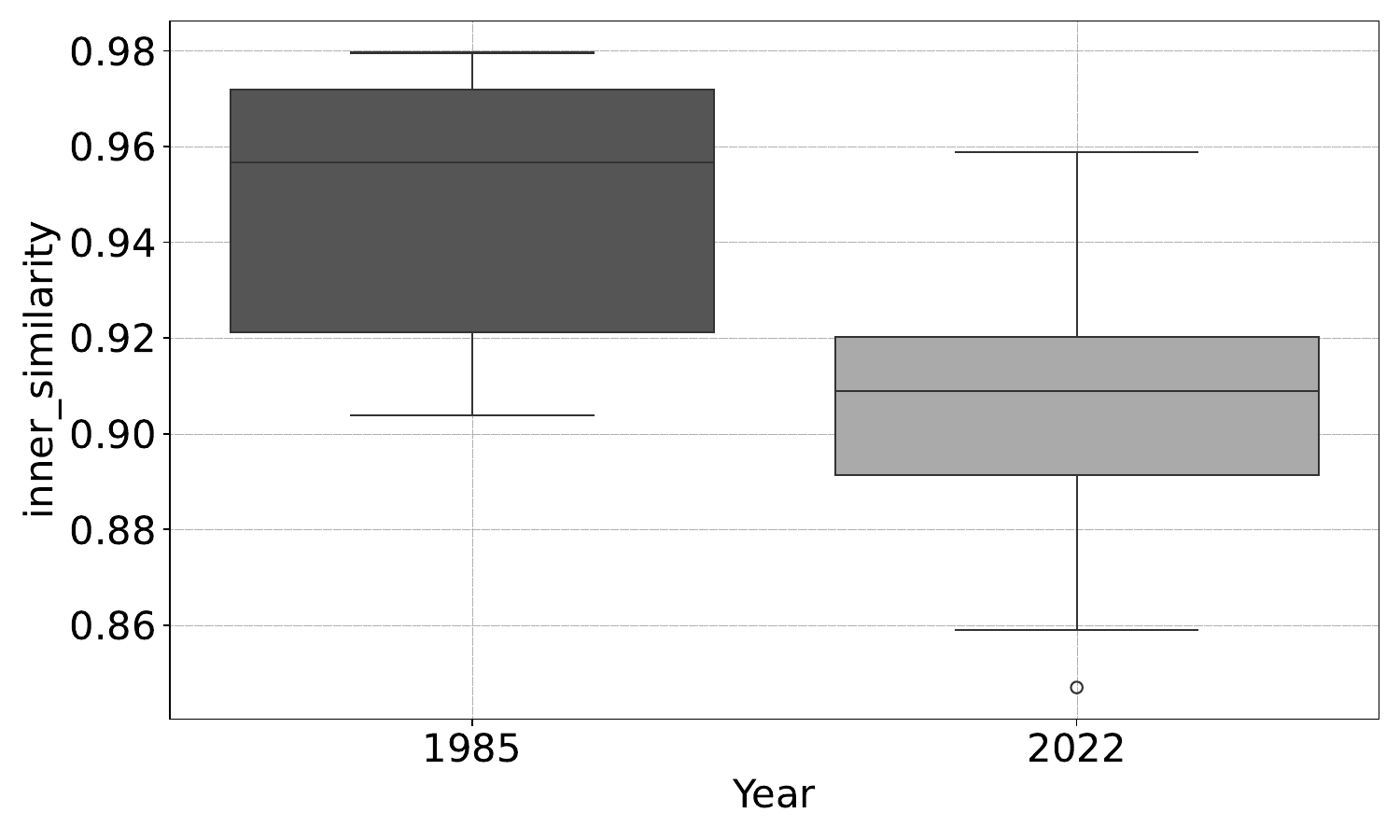}
\caption{Distribution of semantic similarity between the centroids of all Non-AI-engaged and all AI-engaged papers (e.g., \texttt{inner\_similarity}) within fields decreased in 2022 relative to 1985. Values closer to 1 on the y-axis represent higher \texttt{inner\_similarity}.}
\label{fig:inner_similarity}
\end{figure}

\begin{table}[!ht]
\centering

\begin{tabular}{lcccc}
\hline
\textbf{Variable} & \textbf{Count} & \textbf{Mean} & \textbf{Std} \\
\hline
\texttt{ai\_similarity} & 760 & 0.71 & 0.09 \\
\texttt{non-ai\_similarity} & 760 & 0.56 & 0.07 \\
\texttt{inner\_similarity} & 760 & 0.94 & 0.03 \\
\texttt{Ubiquity} & 760 & 0.15 & 0.09 \\
\hline
\end{tabular}
\caption{Summary statistics for variables.
\label{tab:summary_stats}}
\end{table}

Figure~\ref{fig:inner_similarity} shows that overall the inner similarity between a field's AI-engaged and Non-AI-engaged decreased for the period 1985 - 2022, highlighting a widening semantic gap between a field's AI-engaged work and all of its other work. This result holds at the field-level for all but 4 fields in our data. Table~\ref{tab:percentage_change_distances} shows the percentage change by field in the similarity between the AI-engaged papers of Computer Science and the AI-engaged and Non-AI-engaged papers of each field. While many fields saw the similarity between their Non-AI-engaged papers and the AI-engaged work of Computer Science increase, this  increase was, in most cases modest and perhaps due to the increasing interest that Computer Science has taken in disciplinary problems beyond its borders (consider, for instance, the broad and heterogeneous nature of inquiry conducted under the banner of data science and information science).

\begin{table}[ht]
    \centering
    \begin{tabular}{lccc}
        \hline
        \textbf{Field} & \textbf{ai\_sim $\Delta$} & \textbf{non-ai\_sim $\Delta$} & \textbf{inner\_sim $\Delta$}\\
        \hline
        Agricultural \& Food Sciences & 57.9 & 16.0 & -4.0 \\
        Art & -14.1 & -3.0 & 0.1 \\
        Biology & 27.8 & 15.3 & -2.0 \\
        Business & -4.6 & -11.7 & -0.7 \\
        Chemistry & 42.9 & 14.0 & -6.4 \\
        Computer Science & 0.0 & -1.1 & -1.1 \\
        Economics & 19.1 & 1.1 & -3.9 \\
        Education & -12.0 & -17.7 & -1.6 \\
        Engineering & 0.4 & 2.1 & -2.1 \\
        Environmental Science & 42.7 & 9.6 & -12.7 \\
        Geology & 26.6 & 12.8 & -9.1 \\
        History & 12.2 & -9.3 & -7.4 \\
        Linguistics & -9.2 & -9.2 & 1.7 \\
        Materials Science & 42.9 & 5.0 & -9.6 \\
        Mathematics & 10.0 & -9.3 & -7.9 \\
        Medicine & 37.4 & 25.3 & -6.2 \\
        Philosophy & -18.5 & -9.9 & 3.1 \\
        Physics & 28.6 & 10.2 & -5.3 \\
        Political Science & 7.5 & -15.8 & -3.2 \\
        Psychology & 3.7 & 2.1 & 0.6 \\

        \hline
    \end{tabular}
    \caption{Percentage change in distances by field in the period 1985 to 2023}
    \label{tab:percentage_change_distances}
\end{table}

To measure the association between changes in each of our three variables (\texttt{ai\_similarity}, \texttt{non-ai\_similarity}, \texttt{inner\_similarity}), and our dependent variable (\texttt{Ubiquity}), we perform panel regressions with field-level fixed-effects (Equation~\ref{eq:regression}). In particular, we seek to understand how changes in the trajectories of each of our three measured variables are associated with changes in Ubiquity. Specifically, we model this association as:
\begin{equation}
\label{eq:regression}
    \text{\texttt{Ubiquity}}_{it} = \beta_0 + \beta_1 X_{1,it} + \beta_2 X_{2,it} + \beta_3 X_{3,it} + \beta_4 (X_{1,it} \times X_{2,it}) + \alpha_i + \epsilon_{it},
\end{equation}

\noindent
where field-level fixed-effects $\alpha_i$ denote field-specific intercepts and allow us to control for all (stable) differences between fields. (Table~\ref{tab:summary_stats} presents summary statistics for variables used in our regression analysis.) Fixed-effects attempt to account for all unobserved, time-invariant differences across fields and, thereby, controls for any constant characteristics of each field that might influence \texttt{Ubiquity}, ensuring that the estimated relationships are not confounded by these differences. Finally, in this model, $X_{1,it}$, $X_{2,it}$, and $X_{3,it}$ represent \texttt{ai\_similarity}, \texttt{non-ai\_similarity}, and \texttt{inner\_similarity}, respectively, for field $i$ at time $t$. In this model, we interact \texttt{ai\_similarity} and \texttt{non-ai\_similarity} in an effort to evaluate whether the combined influence of these terms on \texttt{Ubiquity} is synergistic (greater than the sum of their individual effects) or antagonistic (less than the sum of their individual effects).

\subsubsection{Findings}
Table~\ref{tab:compact-regression-results} reports the results of our panel regression. We find that increases in the similarity of either a field's AI-engaged papers (\texttt{ai\_similarity}) or a field's Non-AI-engaged papers (\texttt{non\_ai\_similarity}) with the AI-engaged papers of Computer Science are individually associated with increases in the ubiquity of AI engagement (\texttt{Ubiquity}) within the field. That is, when either AI-engaged or Non-AI-engaged papers within a field become more like Computer Science AI-papers, more of the venues within that field participate in publishing AI-engaged papers. However, the strong, negative coefficient on the interaction term indicates that when both \texttt{ai\_similarity} and \texttt{non\_ai\_similarity} increase at the same time, the combined effect on Ubiquity is significantly less than what would be expected from the sum of their individual effects. This result suggests a complex interaction between changes in the semantics of both AI-engaged and Non-AI-engaged papers (\texttt{ai\_similarity} and \texttt{non\_ai\_similarity}) in their relationship with \texttt{Ubiquity}, where the presence of both conditions simultaneously moderates or even reverses increases that would otherwise be observed when each increases independently. Consideration of the association between Ubiquity and \texttt{inner\_similarity} bears this out. When a field's AI-engaged and Non-AI-engaged papers become more similar to one another, field-level ubiquity falls.

This observation complicates the ubiquity story in an interesting and surprising way. Specifically, it suggests that the observed rise in ubiquity is associated with a semantic tension within fields. While AI-engaged papers are, indeed, diffusing across publication venues within fields, this process only accelerates when the semantics of AI-engaged papers and the Non-AI-engaged papers published in the same journals in the same years pull in opposite directions. This result suggests that ubiquity is actually something of an oil and water phenomenon---AI-engaged work is spreading out over fields, but it is not mixing well with Non-AI-engaged work.

\begin{table*}[!ht]
\centering
\begin{tabular}{l|c|c}
\hline
\textbf{Variable} & \textbf{Coefficient} & \textbf{Standard Error} \\
\hline
\texttt{ai\_similarity} & 1.904$^{***}$ & 0.277 \\
\texttt{non-ai\_similarity} & 2.111$^{***}$ & 0.446 \\
\texttt{inner\_similarity} & -0.548$^{***}$ & 0.188 \\
\texttt{ai\_similarity} $\times$ \texttt{non-ai\_similarity} & -3.178$^{***}$ & 0.548\\
\hline
\end{tabular}
\begin{tabular}{l|l|l|l}
Observations: 741 & R$^2$: 0.119 & Adjusted R$^2$: 0.092 & F Statistic: 24.326$^{***}$ (df = 4; 718)
\end{tabular}
\caption{Results of fixed-effects panel regression. $^{***}$ denotes significance at the 0.001 level.}
\label{tab:compact-regression-results}
\end{table*}

\section{Discussion and Concluding Remarks}

While other studies have observed broad increases in AI engagement, this study provides the first comprehensive, empirical analysis of AI's increasing ubiquity within and across diverse scientific and scholastic fields. Given that our study relies on a binary classification technique, we are limited in what we can say about the nature of AI engagement. For instance, it is not clear from our work whether and to what extent the use of AI as part of the practice of science is increasing in ubiquity. In fact, we are unable to distinguish between publications that employ AI, develop novel approaches to AI, or engage with AI from a conceptual standpoint. As a result, our findings are limited to what we can say about engagement with AI in general as opposed to any particular form of engagement. 

Nevertheless, those findings are striking. We observe broad, sharp, and continually increasing engagement with AI across all 20 fields represented in this study. 
Moreover, while we cannot analyze directly the type of AI engagement, we do observe increasing diffusion of AI engagement across publication venues
As publication venues reliably proxy for sub-disciplinary inquiry within fields \cite{boyack2005mapping,rosvall2008maps,merton1973sociology}, this result
suggests that the within-field engagement is heterogeneous in nature (i.e., critical, innovative, fundamental, etc.). 
%Moreover, while we are not able to analyze directly the type of AI engagement, we do observe increasing diffusion of AI engagement across publication venues, suggesting that the within-field engagement is heterogeneous in nature (i.e., critical, innovative, fundamental, etc.). This is because publication venues reliably proxy for sub-disciplinary inquiry within fields. 
Moreover, sub-disciplinary inquiry is not merely picked out by the questions or problems of principal concern, but by the methods and epistemic norms that are marshalled in the service of their address. Finally, while we observe shifts in the semantic character of AI-engaged papers relative to Non-AI-engaged papers and that these shifts are in tension, we do not investigate the underlying `structural' shift in citation papers. It could be that AI-engaged papers \textit{merely} change their semantic character but otherwise engage in scientific investigation in much the same way as Non-AI-engaged papers in the same field. If this were the case, then we would expect the underlying citation distributions to be roughly similar between and across AI-engaged and Non-AI-engaged publications (though prior work \cite{frank2019evolution} suggests an underlying `structural' shift is underway, as well). We plan to address these limitations in future work.

Our findings suggest that AI is no longer concentrated in a few specialized areas but is becoming more integrated into the broader fabric of scientific inquiry. This result perhaps challenges traditional models of technological diffusion, as engagement with the technology appears to be increasing rapidly everywhere at once, perhaps necessitating new theoretical frameworks. Moreover, the observed oil-and-water phenomenon implies that while AI is spreading, it is not seamlessly blending with existing research paradigms.

Policymakers and funding agencies recognize the potential of AI to transform science and society. They should also recognize the widespread but heterogeneous nature of AI engagement across disciplines. Given the observed tension within fields, funding strategies should support not only the development of AI technologies and skills, but also their thoughtful application and critical examination in diverse fields, for instance, through continuous monitoring of AI's impact on different fields. Funders and policymakers could invest in robust data collection and analysis systems to track and evaluate how AI engagement is evolving over time and across disciplines. This work would enable evidence informed decision-making and the timely adjustment of policies to support beneficial developments while addressing emerging challenges, thereby continuing to enhance the benefits of AI while recognizing and actively working to mitigate potential field-specific risks.

\section*{Acknowledgements}
This project benefited greatly from the use of Globus \cite{chard2014efficient} and from the resources of the Division of Research and Faculty Development at Harvard Business School. We acknowledge helpful feedback from audiences at the University of Chicago's Globus Labs, Data Science Institute, and Knowledge Lab, and at Harvard's Digital Data Design Institute.
This work was supported in part by the U.S.\ Department of Energy, Office of Science, under contract DE-AC02-06CH11357. 

\bibliographystyle{alpha}
\bibliography{main}

\end{document}